\documentclass[aps,12pt,superscriptaddress,preprint,floatfix]{revtex4-1}
\usepackage{amssymb,graphics,graphicx}
\usepackage{amsmath,amssymb,amsthm,bm,multirow,longtable}
\renewcommand{\epsilon}{\varepsilon}
\renewcommand{\phi}{\varphi}

\theoremstyle{plain}

\theoremstyle{definition}
\allowdisplaybreaks

\begin{document}
\title{Viable Alternative to the Standard Model}
\author{Paul H. Frampton}
\email{paul.h.frampton@gmail.com}
\affiliation{
Courtyard Hotel, Whalley Avenue, New Haven, CT 06511, USA}
\author{Thomas W. Kephart}
\email{tom.kephart@gmail.com}
\affiliation{Department of Physics and Astronomy, Vanderbilt University, Nashville, TN 37235}
\date{\today}

\begin{abstract}
As an alternative to the standard model, we couple two  scalar doublets such that, in the third family,
one (${\cal H}$) couples only to the top quark, the other ($H$), that is identified as the 126 GeV state already observed,
to the bottom quark and the tau lepton. Three explicit predictions of the model are that the partial decay rate
$\Gamma(H \rightarrow \gamma \gamma)$ is 28.1\% higher than predicted by the standard model, and the two partial decay widths $\Gamma(H \rightarrow \bar{b}b)$ and $\Gamma(H \rightarrow \bar{\tau}\tau)$ are both predicted to be greater than, or equal to, 
their standard model values.

\end{abstract}
\pacs{}
\maketitle

Experimental data from the Large Hadron Collider (LHC),
especially concerning the properties of recently discovered 126 GeV scalar   \cite{Aad:2012tfa,Chatrchyan:2012ufa},
can allow us new insight into the standard model (SM) and the mechanism of electroweak symmetry breaking.
These data are mainly pertinent to the third generation of fermions
in the SM, containing the top ($t$) and bottom ($b$) quarks together
with the tau ($\tau$) lepton. The SM has proven amazingly robust and resilient: we suspect its Achilles heel may 
lie in its Yukawa couplings.

In the SM, the interactions of these three fermions with the unique scalar doublet 
$h$ are characterized by the Yukawa interaction Lagrangian
\begin{equation}
{\cal L}_Y^{(SM)} =  \left[ Y_t ^{(SM)} \bar{t}t  + Y_b^{(SM)}  \bar{b}b + Y_{\tau}^{(SM)} \bar{\tau}\tau \right] h + c.c.
\label{SMYukawa}
\end{equation}
in terms of the quark mass eigenstates. The vacuum expectation value (VEV) of $h$ is give by
\begin{equation}
<h> = V = (\sqrt{2} G_F)^{-1/2} \simeq 246 ~ {\rm GeV}.  
\end{equation}

The exceptionally simple alternative model we consider here has two complex scalar doublets ${\cal H}$ and $H$.
This alternative model has Yukawa interactions according to the Lagrangian
\begin{equation}
{\cal L}_Y^{(Alt)} =  Y_t ^{(Alt)} \bar{t}t {\cal H} + (Y_b^{(Alt)}  \bar{b}b    + Y_{\tau}^{(Alt)} \bar{\tau}\tau )H   + c.c.
\label{AltYukawa}
\end{equation}
where ${\cal H}$ is orthogonal to  the scalar state $H$ discovered in 2012. Since the decays $H \rightarrow \bar{b}b$ and $H \rightarrow\bar{\tau}\tau $
have been seen, we must conclude that $H$ couples directly to $b$ and $\tau$ via Yukawa terms in the Lagrangian. However, $t$ is not presently required to couple to the $H$. The observed decay $H \rightarrow \gamma \gamma$ could have a contribution from a top loop, but that is not required since it would still be visible because it also has a contribution from a $W$ loop. In fact, the rate $\Gamma(H \rightarrow \gamma \gamma)$ is enhanced if the  $H\bar {t}t$ Yukawa coupling is absent, because the top loop tends to cancel the $W$-loop contribution \cite{Ellis:1975ap,Ioffe:1976sd,Shifman:1979eb,Marciano:2011gm}.
Hence we are free to assume that $t$ gets its mass from a second, heavier, Higgs ${\cal H}$ that is potentially a top condensate $<\bar{t}t>$~\cite{Nambu:1961tp,Nambu:1961fr,Nambu:1988bk,Nambu:1990hj}, see also \cite{Miransky:1989ds,Miransky:1988xi,Bardeen:1989ds}.

We assume that the two scalars ${\cal H}$ and $H$ develop VEVs 
\begin{equation}
<{\cal H}> = V_t,  \,\,\,\,\,  <H> = V_b .  
\end{equation}
and define the variables 
\begin{equation}
r_t=\frac{Y_t^{SM}}{Y_t^{Alt}}  = \frac{V_t}{V}
\end{equation}
and
\begin{equation}
r_{\tau}=r_b=\frac{Y_b^{SM}}{Y_b^{Alt}}  = \frac{V_b}{V}.
\end{equation}

To be consistent with the $W$ mass, one has the sum rule \cite{Frampton:2014npa}
\begin{equation}
r_t^2+r_b^2=r_t^2+r_{\tau}^2=1,
\end{equation}
and the usual parameterization is $r_t =\sin \beta$ and $r_b =\cos \beta$.

Clearly $r_t^2, \, \,r_b^2,$  $ r_{\tau}^2 \leq 1$ from the sum rule.
For the two body decays 
$ H \rightarrow \bar{b}b$ and $ H \rightarrow \bar{\tau}\tau $
this leads to two predictions:\\
{\underline{{\it Prediction (i)}}
\begin{equation}
\frac{\Gamma(H \rightarrow \bar{b}b)^{Alt}}{\Gamma(H \rightarrow \bar{b}b)^{SM}}=\frac{1}{r_b^2}\geq 1
\end{equation}
{\underline{{\it Prediction (ii)}}
\begin{equation}
\frac{\Gamma(H \rightarrow \bar{{\tau}}{\tau})^{Alt}}{\Gamma(H \rightarrow \bar{{\tau}}{\tau})^{SM}}=\frac{1}{{r_{\tau}^2}}\geq 1
\end{equation}
As a third prediction, and as mentioned above,  the decay $H \rightarrow \gamma \gamma$ which has  a one loop contribution only from the $W$-boson:

\noindent
{\underline{{\it Prediction (iii)}}
\begin{equation}
 \frac{\Gamma(H \rightarrow \bar{t}t)^{Alt}}{\Gamma(H \rightarrow \bar{t}t)^{SM}} =1.281
\end{equation}
so the rates for two $\gamma$ decay is 28.1\% higher than the standard model, when we follow the calculations in \cite{Ellis:1975ap,Ioffe:1976sd,Shifman:1979eb,Marciano:2011gm}.

Using the LHC data from the CMS collaboration one finds at the three $\sigma$ level that $r_{\tau}^2 \geq 0.581$  (see \cite{Chatrchyan:2014nva,Chatrchyan:2014vua,Grippo:2014zea}) which implies that $\tan \beta \leq 0.85$.

This alternative to the SM is nicely consistent with the model of Nambu \cite{Nambu:1990hj} where his preferred solution, discussed before the top quark was discovered, was $M_t  > 120 $ GeV,  and  $M_h > 200$ GeV. Here we would identify the $h$ with the ${\cal H}$ and are led therefore, following Nambu's guidance, from the BCS model and bootstrap ideas to  expect that $M_{\cal H}$ is approximately given by $M_{\cal H}\sim 2M_t=346.6$ GeV. These ideas are related to the top quark condensate and have been explored in great detail by Nambu. If the $<{\cal H}>$ is to regarded as a $<\bar{t}t>$ condensate, then this condensation can, in turn, lead to the familiar symmetry breaking via $<H>$. This sort of tumbling
relationship \cite{Raby:1979my} can be seen by studying the diagram of a $t/ b$ fermion loop where two ${\cal H}$ fields are attached as vacuum tadpoles to $t$ and two on-shell $H$ fields are attached to $b$. This fermion loop diagram can then provide the negative mass squared of $H$ which triggers the part of the symmetry breaking due to the $H$ developing a vacuum value.

At this stage we remain agnostic about how the Higgs fields ${\cal H}$ and $H$ couple to the first two generations and so we do not yet fix the 2HDM model classification type \cite{Branco:2011iw}. More data, in particular the measurement of $H \rightarrow \bar{c}c$, could prove invaluable in extending the model to
lighter fermions.\\

\noindent
Acknowledgment: The work of TWK was supported by DoE grant\# DE-SC0011981.\\

\end{document}